*Review Manuscript*

# The Future of Data Analysis in the Neurosciences

Danilo Bzdok[1,2,3,4,*] and B. T. Thomas Yeo[5,6,7,8,9]

1 Department of Psychiatry, Psychotherapy and Psychosomatics, RWTH Aachen University, 52072 Aachen, Germany
2 JARA-BRAIN, Jülich-Aachen Research Alliance, Germany
3 IRTG2150 - International Research Training Group
4 Parietal team, INRIA, Neurospin, bat 145, CEA Saclay, 91191 Gif-sur-Yvette, France

[5] Department of Electrical and Computer Engineering, National University of Singapore, Singapore 119077

[6] Clinical Imaging Research Centre, National University of Singapore, Singapore 117599

[7] Singapore Institute for Neurotechnology, National University of Singapore, Singapore 117456

[8] Memory Networks Programme, National University of Singapore, Singapore 119077

[9] Martinos Center for Biomedical Imaging, Massachusetts General Hospital, Harvard Medical School, Charlestown, MA 02129, USA

* Correspondence: danilo.bzdok@rwth-aachen.de (D. Bzdok)

Prof. Dr. Dr. Danilo Bzdok

Department of Psychiatry, Psychotherapy and Psychosomatics

Pauwelsstraße 30

52074 Aachen

Germany

phone: +49 241 8085729

**Abstract**

Neuroscience is undergoing faster changes than ever before. Over 100 years our field qualitatively described and invasively manipulated single or few organisms to gain anatomical, physiological, and pharmacological insights. In the last 10 years neuroscience spawned quantitative big-sample datasets on microanatomy, synaptic connections, optogenetic brain-behavior assays, and high-level cognition. While growing data availability and information granularity have been amply discussed, we direct attention to a routinely neglected question: How will the unprecedented data richness shape data analysis practices? Statistical reasoning is becoming more central to distill neurobiological knowledge from healthy and pathological brain recordings. We believe that large-scale data analysis will use more models that are non-parametric, generative, mixing frequentist and Bayesian aspects, and grounded in different statistical inferences.

**Introduction**

During most of neuroscience history, new insights were "inferred" with little or no reliance on statistics. Qualitative, often anecdotal reports have documented impairments after brain lesion [1], microscopical inspection of stained tissue [2], electrical stimulation during neurosurgery [3], targeted pharmacological intervention [4], and brain connections using neuron-transportable dyes [5]. Connectivity analysis by axonal tracing studies in monkeys exemplifies biologically justified "inference" with many discoveries since the 60s [6]. A colored tracer substance is injected in vivo into source region A, uptaken by local neuronal receptors, and automatically transported in axons to target region B. This observation in *a single monkey* allows *extrapolating* a *monosynaptical connection* between region A and B to the *entire monkey species* [7]. Instead, later brain-imaging technology propelled the data-intensive characterization of the mammalian brain and today readily quantifies axonal connections, myeloarchitectonic distributions, cytoarchitectonic borders, neurotransmitter receptors, and oscillatory coupling [8-11]. *Following many new opportunities to generate digitized brain data, uncertainties about neurobiological phenomena henceforth required assessment in the statistical arena.*

In the quantitative sciences, the invention and application of statistical tools has always been dictated by changing contexts and domain questions [12]. As this appears to be less appreciated in current neuroscience, the present paper reviews how "optimal" statistical choices are likely to change due to the progressively increasing granularity of digitized brain data. Massive data collection is a game changer in neuroscience [9, 13], and many other public and scientific areas [14-16]. There is an always-bigger interest in and pressure for data sharing, open access, and building "big-data" repositories [8, 17, 18]. For instance, UK Biobank is a longitudinal population study dedicated to the genetic and environmental influence on mental and other disorders. 500,000 enrolled volunteers undergo an impressive battery of clinical diagnostics from brain scans to bone density with a >25 year follow-up. Targeted reanalysis of such national and international data collections will soon become the new normal in neuroscience and medicine. Statistical scalability to high-dimensional data will therefore be inspected from the independent perspectives of i) parametric versus non-parametric, ii) discriminative versus generative, and iii) frequentist versus Bayesian statistical models as well as iv) classical hypothesis testing and out-of-sample generalization.

**Towards adaptive models?**

Today, *parametric* models are still the obvious choice in neuroscience. For instance, many big-sample studies (i.e., data from hundreds of animals or humans) currently apply the same parametric models as previous small-sample studies (i.e., a few dozen animals or humans). However, concentrating on sample size, parametric analyses such as Student's t-test, ANOVA, and Pearson's linear correlation on brain data from many hundred individuals may not yield an improved quality of statistical insight on a neurobiological phenomenon that could not already be achieved with a dozen participants (Box 1). An important caveat relies in their systematic inability to grow in complexity no matter how much data is collected and analyzed [19]. In any classification, a linear *parametric* classifier will always make predictions based on a linear decision boundary between classes, whereas a *non-parametric* classifier can learn a non-linear boundary *whose shape grows more complex with more data*. Analogously, parametric independent component analysis (ICA) and principal component analysis (PCA) require presetting the component number, yet the form and number of clusters in brain data is not known to neuroscientists. Such *finite mixture models* may give way to *infinite mixture models* that reframe the cluster number as a function of data availability and hence gradually yield more clusters without bound. Similarly, classical hidden Markov models may get upgraded to infinite hidden Markov models and we might see more applications of non-parametric decision trees and nearest neighbors in future neuroscientific studies. Finally, it is currently debated whether increasingly used "deep" neural network algorithms with many non-linear hidden layers are more accurately viewed as parametric or non-parametric.

A crucial statistical property for data-rich neuroscience will be *the strength of non-parametric models to automatically increase the number of model parameters*. Many interesting phenomena in the brain are likely to be very complex. Fortunately, probability theory recently proposed stochastic processes that instantiate random variables ranging over unlimited function spaces. *Gaussian Processes* (GP) are an important non-parametric example for specifying distributions on *unknown functions* to place minimal structural assumptions or constraints on possible interactions [20]. Although GPs instantiate infinite dimensional objects, indexing a subset of random variables in this infinite collection is ensured to yield multivariate Gaussian distributions. Instead of fitting one parameter to each variable to predict a behavior or clinical outcome with linear regression, GPs learn a non-parametric distribution of non-linear functions to explain brain-behavior associations. This is likely to leverage predictive regression and classification performance in large-sample studies in neuroscience. For instance, natural scaling to the high-dimensional scenario was demonstrated by a GP regression model that explained 70% of known missing heritable variability in yeast phenotypes [21]. Such emergent insight from complex non-additive interactions between gene loci is difficult to infer by genome-wide association studies (GWAS) that use parametric models to explain small fractions of total heritable variation. Unfortunately, the computational cost of parametric alternatives scales exponentially with interaction order and the necessary correction for the multiple tested hypotheses becomes challenging.

As another non-parametric family, *kernel*-based models can provide statistical advantages by mapping brain variables to a latent variable space [22]. Kernels promise effective modality fusion to incorporate heterogeneous sources given that, mathematically, kernel addition equates with concatenation of variable spaces. Such *genuine multi-modal analysis* can perform conjoint inference on behavioral indices, brain connectivity and function phenotypes. Non-parametric prediction in classification and regression is thus performed based on *dot products on the vectors of the inner product space* without requiring an explicit mapping from the actual brain variables (i.e., "kernel trick"). In this virtual variable space the added dimensions enable linear separability of complex neurobiological effects that are not linearly separable in the original variables. Kernelization of statistical estimators inherits enriched transformation of the input data that only grow linearly in dimension with increasing sample size. Such purposeful increase of input dimensionality and model complexity is useful for small to intermediate, but probably not extremely large datasets because the kernel matrix can grow to terabytes scales. Disadvantages of kernels include the inability to assess contributions of individual variables and to distinguish informative and noise variables. Finally, kernel-based statistics relate to psychology in that learning pairwise similarity measures between environmental stimuli is an aspect of intelligent behavior. Radial basis kernels were argued to offer psychological and neurobiological plausibility as computational model for human decision-making [23].

In sum, neurobiology is high-dimensional in nature and thus difficult to understand for human intuition. By expressing brain phenomena in some finite-dimensional space, parametric models are more interpretable, easier to implement, and faster to estimate. They are often the best choice in data scarcity, yet they are further away from neurobiological ground truth. Exclusive reliance on parametric analysis of brain recordings may keep neuroscientists from extracting structured knowledge on *emergence properties* that can only be grasped in data richness [15, 19, 24]. Even if more complex models do not always result in greater insight [25], non-parametric approaches are naturally prepared to capture complex relationships. This is because the complexity of statistical structure and hence extracted neurobiological insight grow boundlessly with the amount of fitted data [19, 26]. This is likely to be important for modern neuroscience and personalized medicine.

**Towards models that incorporate biological structure?**

The fact that more interpretable models typically require more data samples explains why, in neuroscience, the less data-hungry *discriminative* models have been ubiquitous in small-sample studies, while the popularity of more interpretable *generative* models is steadily increasing with recent data availability (Box 2). For instance, hidden Markov models have recently been applied to high-dimensional temporal data of magnetencephalographic recordings [27]. These generative models simultaneously inferred the spatial topography of the major brain networks subserving environment responses and their cross-talk dynamics without anatomical assumptions. The model-immanent estimation of the joint distributions between the voxel inputs has identified 100-200ms windows of coherent spatiotemporally states and their exact functional coupling occurring faster than previously thought (i.e., both are hidden variables). As another example, neuroscientists often conceptualize behavioral tasks as recruiting multiple neural processes supported by multiple brain regions. This century-old notion [28] was lacking a formal mathematical model. The conceptual premise was recently encoded with a generative model [29]. Applying the model to 10,449 experiments across 83 behavioral tasks revealed heterogeneity in the degree of functional specialization within association cortices to execute diverse tasks by flexible brain regions integration across specialized networks [29, 30]. As a clinical example, dynamic causal modeling for model-induced variable spaces and ensuing group classification by learning algorithms were combined to infer neurobiologically interpretable manifolds underlying aphasia [31]. Their model identified the directed interregional connectivity strengths that distinguish aphasic and neurotypical individuals. This generative modeling approach was more explanatory than brute-force prediction based on activity extent or linear correlations. The influence exerted from right planum temporale and right Heschl's gyrus on their left counterparts turned out as candidate mechanisms underlying communication impairment and formally discarded other plausible auditory processing schemes. Such generative approaches may refine diagnosis and treatment of spectrum disorders in neurology and psychiatry.

Moreover, generative approaches to fitting biological data have successfully reverse-engineered i) human facial variation related to gender and ethnicity based on genetic information alone [32], ii) the complexity gradient in the ventral-visual processing hierarchy which confirmed encoding of low-/mid-/high-level facets in V1, V2, and V4 mostly known from animals [33], iii) structural content from naturalistic static images by revealing an explicit inverse mapping between occipital activity and retinotopic stimulation [34], iv) video scenes from naturalistic movies enabled by revealing importance of separately modeling slow and fast visual movement [35], and v) brain atlases of semantic specificity from naturalistic speech reinforcing the intuition that various different semantic features are widely distributed in the cortex [36]. Finally, it is becoming increasingly clear that neural-network models constitute a class of computational architectures [37] that lend themselves particularly well to representation and manifold learning tasks [38]. In particular, autoencoders are modern neural-network models that have been formally shown to generalize commonly employed representation discovery techniques, such as ICA and PCA as well as clustering models [39, 40]. Autoencoder architectures therefore have the generative potential to extract local (brain-area-like) and global (brain-network-like) representation to abandon hand-crafted "feature engineering" made difficult by missing ground truth [41]. Neurobiologically valid representations are revealed as sets of predictive patterns that together explain psychological tasks and disease processes without being constrained to functional specialization into distinct regions or functional integration by brain networks. The predictive validity of the discovered functional compartments thus extends the interpretational spectrum offered by explained-variance approaches like ICA and PCA.

In sum, neuroscientists can afford more generative models as brain data become abundant. In contrast to discriminative models, their naturally higher interpretability could open the black box of the neural processing architectures underlying behavior and its disturbances. Generative models enable *more detailed understanding* by exposing the low-dimensional manifolds embedded within high-dimensional brain data. However, "the more detailed and biologically realistic a model, the greater the challenges of parameter estimation and the danger of overfitting" [42]. A crucial next step in neuroscience might lie in extracting actual pathophysiological *mechanisms* from brain measurements in mental disorders. In contrast, applying discriminative approaches on patients grouped by the diagnostic manuals DSM or ICD could recapitulate categories that are neither neurobiologically valid nor clinically predictive [43]. Finally, discriminative models may be less potent to characterize the neural mechanisms of information processing up to the ultimate goal of recovering subjective mental experience from brain recordings [31, 44-46].

**Towards integration of traditional modeling regimes?**

After the 19th century was driven by *Bayesian* and the 20th century by *frequentist* statistics [47], one might wonder about their relative contributions in the 21st century. These two attitudes towards quantitative investigation are the most common distinction in statistics in general [48] and in neuroscience in particular [49, 50], yet orthogonal to the parametric/non-parametric and discriminative/generative perspectives on statistical models [48]. Overall, the many desirable properties of Bayesian modeling and interpretation are contrasted by the computationally feasibility of frequentist models in high-dimensional data (Box 3). In neuroscience the speed-accuracy tradeoff in Bayesian posterior inference has been advantageously rebalanced using variational Bayes for limited computational budgets on a number of previous occasions. This includes Bayesian time-series analysis [51], model selection for group analysis [50] and mixed-effects classification for imbalanced groups [52]. Bayesian inference is an appealing framework by its intimate relationship to properties of firing in neuronal populations [53] and the learning human mind [44]. Additionally, it is important for currently increasing efforts in *transdiagnostic* clinical neuroscience [54] that Bayesian hierarchical models could elegantly handle *class imbalances* with very unequal group sizes. Moreover, *inferring hierarchies of statistical relationships*, a natural way of inducing parsimony, is difficult or impossible in a frequentist model selection regime and much more feasible in a Bayesian model averaging regime.

In particular, recent advances in *non-parametric Bayesian* methods [55] combined with extensive datasets promise progress in longstanding problems in cognitive and clinical neuroscience. As a key problem in cognition, neuroscientists have not agreed on a description system of mental operations (called 'taxonomy' or 'ontology') that would canonically motivate and operationalize their experiments [56, 57]. As a key problem in medicine, partly shared neurobiological endophenotypes are today believed to contribute to the pathophysiology of various neurological and psychiatric diagnoses (called 'nosology') despite drastically different clinical exophenotypes [31, 43, 58]. As an interesting observation, both these neuroscientific challenges can be statistically recast as *latent factor problems* [cf. 59]. The same class of statistical models could both identify the unnamed building blocks underlying human cognition and the unknown neurobiological mechanisms underlying diverse mental disorders. For instance, hierarchical Bayesian models were recently borrowed from the topic modeling domain to estimate a latent cognitive ontology [60] and neurobiological subtypes in Alzheimer's disease [61]. Further, formal posterior inference in non-parametric Bayesian models could gracefully handle complexity in the brain by estimating the *number* of latent factors in cognition and disease using Chinese Restaurant Processes [62], *relative implications* of latent causes in neurobiological observations using Indian Buffet Process [63], as well as induce *hierarchies* of cognitive primitives and disease endophenotypes using Hierarchical Dirichlet Processes [64]. Cluster detection in the non-parametric Bayesian regime exchanges the neurobiologically implausible winner-takes-all property of traditional clustering by allowing each observation to participate in *all clusters*. Crucially, the non-parametric aspect of these Dirichlet Process models allows the number of inferred clusters to grow organically with increasing sample size.

In sum, the scalability of model estimation in the data-rich scenario is calibrated between frequentist numerical optimization and Bayesian numerical integration. Ingredients from both statistical regimes can be advantageously combined by adjusting the modeling goal [65, chapter 5, 66]. Many frequentist methods reliant on gradient-based optimization can be recast as Bayesian integration problems, for instance using Hamiltonian MCMC methods. Conversely, integration problems can be turned into optimization problems, for instance using variational Bayes approximations. High-dimensional data have been authoritatively argued to motivate novel blends between simplier-to-use frequentist and more integrative Bayesian modeling aspects [47]. Indeed, the continuously increasing number of observations and number of parameters to estimate have earlier been predicted to complicate logical consistency in statistical theory [67]. We therefore believe that the recent emergence of extensive datasets in neuroscience could analogously prompt emergence of more frequentist-Bayesian hybrid approaches.

**Towards diversification of statistical inference?**

Drawing statistical inference on regional brain responses during controlled experiments has largely hinged on *classical null-hypothesis rejection*, but is increasingly flanked by *out-of-sample prediction* based on *cross-validation* [68, 69]. Both inferential regimes provide formal justifications for deriving neurobiological knowledge using mathematical models (Box 4). *Classical inference* measures the *statistical significance* associated with a relationship between typically few variables given a prespecified model. *Generalization inference* assesses the *robustness of patterns* between typically many variables in the data [70]. One might think that these differences in deriving formal conclusions on brain recordings are mostly of technical relevance but there is an often-overlooked misconception that models with high *explanatory power* always exhibit high *predictive power* [71, 72]. A neurobiological effect assessed to be statistically significant by a p-value may sometimes not yield successful predictability based on cross-validation, and vice versa. Their theoretical differences are also practically manifested in the high-dimensional setting where classical inference needs to address the *multiple comparisons problem* and generalization inference involves tackling the *curse of dimensionality* [73, 74].

Consequently, care needs to be taken when combing both inferential regimes in data analysis [68, 69]. Say, a neuroscientist wants to predict Alzheimer diagnosis from the >100,000 brain location per brain scan by sparse L1-penalized logistic regression (i.e., classification with minimal necessary variables) using cross-validation but performs preliminary dimensionality reduction to the most important 10,000 voxels by ANOVA-mediated variable ranking using classical inference. In this case, it is not permitted to conduct an ordinary significance test (i.e., classical inference) on the obtained sparse model coefficients (obtained from generalization inference) because it would involve recasting an originally high-dimensional variable-selection as a univariate setting [75]. The ANOVA test would ignore the fact that the sparse logistic regression had already reduced the variables to the most important ones [72]. Very recently proposed methods for so-called *post-selection inference* allow replacing *naive* by *selection-adjusted p-values* for a set of variables previously chosen to be meaningful predictors [76]. This and similar clashes between inferential regimes will probably increase in neuroscientific data analysis.

However, classical inference and generalization inference have also been advantageously joined towards a same neuroscientific goal. In a first example, cross-validated classification algorithms estimated the relative contribution of all macroscopical brain networks in a battery of psychological tasks, while ANOVA was used to find smaller subsets of most important networks for each task [45]. In a second example, Latent Dirichlet Allocation [77] was used to find a nested hierarchy of volume atrophy endophenotypes in Alzheimer's disease, while these were cross-validated by distinct trajectories in memory and executive function decline using classical hypothesis testing [61]. More generally, a neurobiological question, such as "Are region A and B significantly connected?", can also be cross-validated by an independent method that can make the same observation, such as significant structural and functional connectivities [78]. The previous two examples performed cross-validation in a *different kind of data*, in contrast to the typical practice of performing cross-validation in *unseen data of the same kind*. This is important because fMRI, EEG, MEG, fNIRS, and other neuroscientific methods measure biological phenomena only indirectly. Out-of-sample generalization to a different modality (e.g., behavior, genetics, microbiomics) increases confidence that the findings reflect neurobiological reality. Combining different inferential regimes can therefore corroborate neurobiological findings.

In sum, the leap from quantitative brain measurements to neurobiological knowledge is secured by statistical inference. There is not one but several different types of inference that can ask a same question with different mathematical foundations requiring differently nuanced neuroscientific interpretation. Historically, classical inference was invented for problems with small samples that can be addressed by plausible, handpicked models with a small number of parameters [cf. 79]. Some authors therefore emphasize that "one should *never* use sum of squared errors, p-values, $R^2$ statistics, or other classical measures of model fit on the training data as evidence of a good model fit in the high-dimensional setting." [80, p. 247, their emphasis]. P-values and other classical guarantees may lose their ability to evaluate model fit in high-dimensional data. Instead, out-of-sample generalization by *successful cross-validation to independent data* will be increasingly used given natural tuning to problems with more variables and larger samples [cf. 24]. Envisioning a future of precision medicine, only cross-validated predictive models can obtain answers from a single data point [42]. Finally, data richness will increasingly require preliminary dimensionality-reduction and feature-engineering procedures, including k-means clustering and ICA decomposition, that do not themselves perform statistical inference. A back and forth between dimensionality-reducing data transformations, inductive pattern extrapolation and deductive hypothesis testing of the discovered candidate effects will become valid routines in neuroscience.

**Concluding Remarks and Future Perspectives**

After astronomy, particle physics, and genetics, massive data is the new reality in neuroscience and medicine. Rich datasets can extend the spectrum of possible findings and permissible conclusions about the brain. Yet, the unavoidable impact on data analysis practices is currently shunned. While sample size and information granularity are increasing progressively, the prompted shift in statistical choices may be categorical. Neuroscientists need to extend their modeling instincts towards neurobiological insight with gradually increasing quality of insight as data accumulate [12] and towards prediction on the single-individual level [81]. It is difficult to overstate the importance of closing training gaps in next-generation PhD curricula that will include machine-learning, computer programming, distributed multi-core processing, and advanced visualization [82]. In a nutshell, neuroscience is entering the era of large-scale data collection, curation, and collaboration with a pressing need for statistical models tailored for high-dimensional inference. These will frequently lie beyond the scope of the statistical repertoire cherished today. Analyzing extensive multi-modal datasets with inappropriate statistical models would be a waste of public financial resources and our limited scientific efforts.

**TEXT BOX 1: Parametric and non-parametric models**

This statistical distinction takes place at the equilibrium between imposing specific assumptions (i.e., *parametric*) and not assuming a certain functional form (i.e., *non-parametric*) to encourage discovery of relevant structure driven by the brain data themselves [19, 83]. Contrary to common misunderstanding, both classes of quantitative modeling involve parameters. "Non-parametric" is defined in three different flavors: The first, perhaps most widespread meaning implies those statistical estimators that do not make explicit assumptions about the *probability distribution* (e.g., Gaussian) from which the data have arisen. As a second, more intuitive explanation, non-parametric models do not assume that the *structure* of the statistical model is fixed. The last, most formal definition emphasizes that *the number of model parameters* increases explicitly or implicitly with the number of available data points (e.g., number of brain recordings). In contrast, the number of model parameters are fixed in parametric models and do not vary with sample size (Fig. 1). In its most extreme manifestation, effectively fitting an arbitrary functional form can result in non-parametric models larger in memory usage than the actual input data themselves. Flexible non-parametric models include decision trees like random forests, nearest-neighbor estimators, kernel support vector machines or kernel principal component analysis, and hierarchical clustering, as well as bootstrapping and other resampling procedures. More rigid parametric models include methods from classical statistics based on maximum likelihood estimation, Gaussian mixture models, support vector machines, principal component analysis, and k-means clustering, but also modern penalized regression variants like Lasso, ElasticNet, and RidgeRegression.

It is an advantage of parametric modeling to expresses the data compactly in a few model parameters, which increases interpretability, requires smaller datasets, has higher statistical power, and incurs lower computational load. Additionally, if the parametric assumptions hold "true" in nature, non-parametric approaches can be less powerful than their parametric counterparts. Although parametric and non-parametric models require assumptions, non-parametric models have the advantage to be more robust about estimating properties of still poorly known phenomena in the brain if the modeling assumptions are partly false. Due to their robustness, non-parametric methods may prevent improper use and misunderstanding. Most importantly, even if the number of parameters in parametric models can be *manually* increased, only non-parametric models have an inherent ability to *automatically* scale with the complexity of the available data resources.

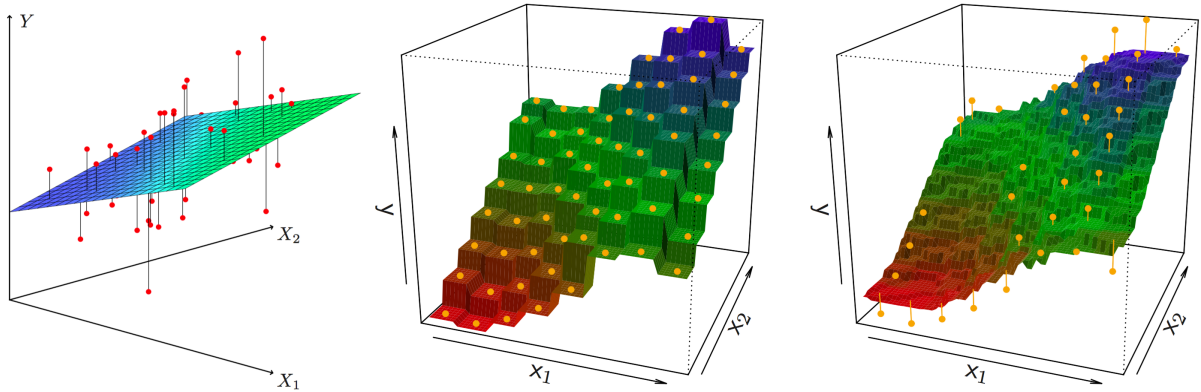

**Figure 1 (in Box 1): Prediction based on parametric versus non-parametric regression**

Fitted models that predict the continuous outcome Y based on the observed variables $X_1$ and $X_2$. *Left*: Ordinary linear regression finds the best plane to distinguish high and low outcomes Y. *Middle/Right*: K-nearest neighbor regression predicts the same outcome Y based on the K=1 (*middle*) or K=9 (*right*) closest data points in the available sample. Parametric linear regression cannot grow more complex than a stiff plane (or hyperplane when more input dimensions $X_n$) as decision boundary, which entails big regions with identical predictions Y. Non-parametric nearest-neighbor regression can grow from a rough step-function decision boundary (k=1) to an always-smoother fit (k=9) by incorporating the data in more complex ways. Non-parametric models therefore outperform parametric alternatives in many data-rich scenarios. Reused with permission from [80].

**TEXT BOX 2: Discriminative and generative models**

*Discriminative* models are common choices when best-possible prediction of neurobiological phenomena is the primary aim (e.g., based on behavioral phenotype, age, performance or clinical scores), which does not require recovery of the explicit relationships between input variables (Fig. 2). *Generative* models also aim at successful extrapolation of statistical relationships to new data but simultaneously extract special neurobiological structure from brain data by *indirectly inferring the target variable from hidden variables* [85]. Typical discriminative models include linear regression, support vector machines, decision-tree algorithms, and logistic regression, while generative models include hidden Markov models, modern neural network algorithms, dictionary learning methods, and many non-parametric statistical models [77]. As an important advantage in brain science, generative modeling approaches are more expressive in extracting representations in input data that are "economical to describe but allow the input to be reconstructed accurately" [86].

More formally, discriminative models focus on a target variable y, whereas generative models are concerned with the *joint distributions* between observed and unobserved variables [84]. Discriminative models find a *direct* mapping function by predicting a categorical or continuous target variable y by estimating P(y|x) without capturing special properties in the data x. Generative models estimate relevant properties in the data by P(y|x) from P(x|y) and P(y). Generative models can hence produce synthetic, never observed examples x~ for targets y by sampling from the estimated joint distributions P(x,y). The strength of generative models to conjointly realize predictive modeling and representation learning is paid by requiring more input data, more computational resources, and possibly more model parameters to fit. Consequently, with too few data points available for model estimation, discriminative models will usually achieve better prediction performance in new data because generative models suffer from higher model variance. Finally, generative models make an assumption about the distribution of the input data and their performance therefore suffers more if this model is inaccurate. Discriminative models are generally more robust to assumptions about the data-generating process.

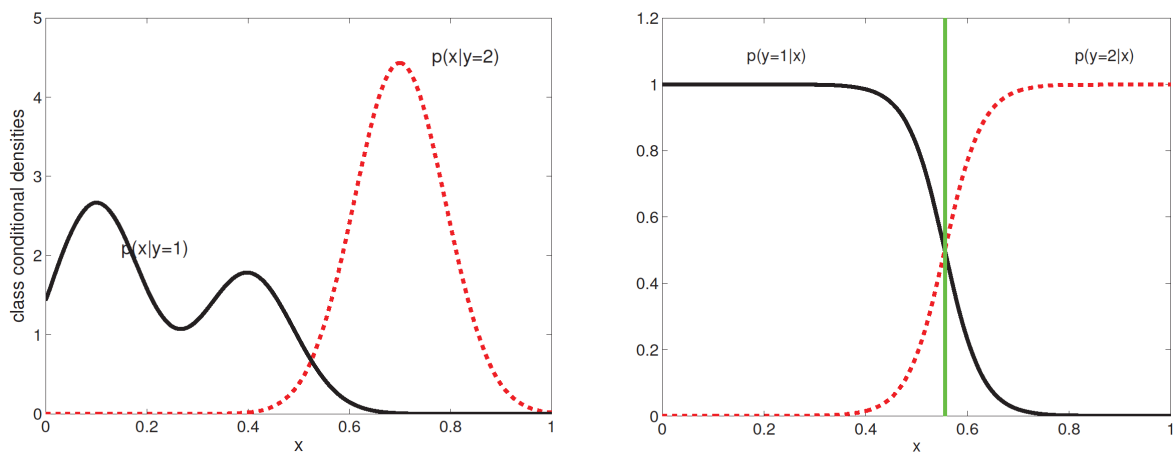

**Figure 2 (in Box 2): Class-conditional densities can be more complex than class posteriors**

Generative models (*left*) estimate P(x|y=c) by indirectly fitting mechanisms in data conditioned on the class c. Discriminative models (*right*) estimate the posterior probability P(y=c|x) to directly quantify the presence of class c conditioned on the data x without capturing data mechanisms underlying that class. It is important to note that peaks of the class-conditional densities P(x|y=c) have no impact on the class posteriors P(y=c|x). Only generative models can therefore produce new, unseen examples x~ from the captured input distributions characteristic for class c. Figure reused with permission from [65].

**TEXT BOX 3: Frequentist and Bayesian models**

In theory, the *frequentist* attitude aims at universally acceptable, investigator-independent conclusions on neurobiological processes by avoiding hand-selected priors on model parameters. The *Bayesian* attitude is more transparent in the unavoidable, necessarily subjective introduction of existing domain knowledge as model priors. Frequentist modeling achieves best-guess values (often without uncertainty interval) treating the model parameters as fixed unknown constants and averaging over input data treated as random. Instead, ideal Bayesian modeling achieves full posterior distribution estimates (including intervals) by averaging over the random model parameters rather than the data. Frequentist approaches typically perform *model selection* to obtain model parameters by maximum likelihood estimation using numerical optimization. Bayesian approaches typically perform *model averaging* by integrating over model parameters conditioned on prespecified priors by approximate resampling solutions from Markov chain Monte Carlo (MCMC) variants.

In practice, statistical models span a continuum between the extreme poles of frequentism and Bayesianism. They exhibit many unexpected similarities [65, 83]. For instance, the *bootstrap* is a frequentist method for population-level inference of confidence intervals and non-parametric null-hypothesis testing [87] that also lends itself to Bayesian interpretations [70]. Frequentist practitioners are mostly concerned about model overfitting, while Bayesian practitioners are more concerned about the pertinence and number of domain-informed priors.

Important for data-intensive brain science, the frequentist-Bayesian tradeoff impacts the computational budget required for model estimation (Fig. 3). Generally, the more one adheres to frequentist instead of Bayesian ideology, the less computationally expensive and the less technically involved are the statistical analyses. It is a widespread opinion that Bayesian models do not scale well to the data-rich setting, although there is little work on the behavior of Bayesian methods in high dimensions [88, 89]. While the pure frequentist approach computes maximum likelihood estimation, the pure Bayesian approach achieves *inference* by full posterior probability distributions computed by *asymptotically exact* MCMC. Given their computational cost, MCMCs have mainly been used for small-scale problems. The practical applicability of Bayesian methods has been greatly enhanced through the development of *approximate inference algorithms* such as variational Bayes and expectation propagation [90]. Nevertheless, exact MCMC and its accelerated approximate variants tend to suffer from i) uncertainty when convergence is reached, ii) difficult-to-control "random-walk" behavior, and iii) limited scaling to the high-dimensional setting [91]. Consequently, the intractability of Bayesian posterior integrals motivated a rich spectrum of model hybrids [47]. There is an increasing trend towards *incorporating appealing Bayesian aspects into computationally cheap frequentist approaches* [e.g., 92].

| Method | Definition |
|---|---|
| Maximum likelihood | $\hat{\boldsymbol{\theta}} = \mathrm{argmax}_{\boldsymbol{\theta}}\, p(\mathcal{D}|\boldsymbol{\theta})$ |
| MAP estimation | $\hat{\boldsymbol{\theta}} = \mathrm{argmax}_{\boldsymbol{\theta}}\, p(\mathcal{D}|\boldsymbol{\theta})p(\boldsymbol{\theta}|\boldsymbol{\eta})$ |
| ML-II (Empirical Bayes) | $\hat{\boldsymbol{\eta}} = \mathrm{argmax}_{\boldsymbol{\eta}} \int p(\mathcal{D}|\boldsymbol{\theta})p(\boldsymbol{\theta}|\boldsymbol{\eta})d\boldsymbol{\theta} = \mathrm{argmax}_{\boldsymbol{\eta}}\, p(\mathcal{D}|\boldsymbol{\eta})$ |
| MAP-II | $\hat{\boldsymbol{\eta}} = \mathrm{argmax}_{\boldsymbol{\eta}} \int p(\mathcal{D}|\boldsymbol{\theta})p(\boldsymbol{\theta}|\boldsymbol{\eta})p(\boldsymbol{\eta})d\boldsymbol{\theta} = \mathrm{argmax}_{\boldsymbol{\eta}}\, p(\mathcal{D}|\boldsymbol{\eta})p(\boldsymbol{\eta})$ |
| Full Bayes | $p(\boldsymbol{\theta}, \boldsymbol{\eta}|\mathcal{D}) \propto p(\mathcal{D}|\boldsymbol{\theta})p(\boldsymbol{\theta}|\boldsymbol{\eta})p(\boldsymbol{\eta})$ |

**Figure 3 (in Box 3): Different shades of Bayesian inference**

There is no unique Bayesian formulation to perform statistical inference, but there is a variety of them. For instance, *type-II maximum likelihood* or *empirical Bayes* has genuine frequentist properties, does not specify a prior distribution before visiting the data, and is often used in non-Bayesian modeling. Generally, the more integrals to be solved or approximated, the higher the computational budget needed for model estimation. Reused with permission from [65].

**BOX 4: Null-hypothesis testing and out-of-sample generalization**

Statistical inference can be generally defined as the extraction of new knowledge from parameters in mathematical models fitted to data [15]. In *classical inference,* invented almost 100 years ago [93], the neuroscientist articulates two mutually exclusive hypotheses by domain-informed judgment with the agenda to disprove the null hypothesis embraced by the research community. A *p-value* is then computed that denotes the conditional probability of obtaining an equal or more extreme test statistic provided that the null hypothesis is correct at the conventional significance threshold alpha=0.05 [94]. For instance, statistical significance of the Student's t-test indicates a difference between two means with a 5% chance after sampling twice from the same population. Adopting *deductive reasoning* and a mostly *retrospective* viewpoint [95], state-of-the-art hypotheses are continuously replaced by always more pertinent hypotheses using *verification* and *falsification* in a Darwinian process. The classical framework of null-hypothesis falsification to infer new knowledge is still the go-to choice in many branches of neuroscience. Considering the data-rich scenario, it is an important problem that p-values intrinsically become better (i.e., lower) as the sample size increases [96]. In some fields, it is therefore now mandatory to report effect sizes in addition or instead of p-values [94].

In contrast, *generalization inference* emerged much more recently in basic statistics [97] and tests whether complex patterns extrapolate to independent data using *cross-validation*. Formally, this inferential regime is mathematically justified by the so-called *Vapnik-Chervonenkis* (VC) *dimensions* from *statistical learning theory*, conceptually related to the degrees of freedom from classical statistics [98]. In generalization inference on high-dimensional patterns, the VC dimensions formally express the circumstances under which a class of functions is able to learn from a finite amount of data to successfully predict a given neurobiological phenomenon in unseen data [99]. Thus, the VC dimensions provide a probabilistic measure of whether a certain model is able to learn a distinction given a dataset. This *inductive reasoning* to learn a general principle from examples adopting a mostly *prospective* viewpoint contrasts the deductive, retrospective logic of null-hypothesis testing. In practice, cross-validation is frequently used to quantify the out-of-sample performance by an unbiased estimate of a model's capacity to generalize to data samples acquired in the future [100]. Model assessment is done by training on a bigger subset of the available data (i.e., *training set*) and subsequently applying the trained model to the smaller remaining part of data (i.e., *test set*).